\documentstyle[amssymb,preprint,aps]{revtex}
\begin{document}
\draft
\title{Microwave surface impedance of proximity-coupled
superconducting (Nb) / spin-glass (CuMn) bilayers.}
\author{L. V. Mercaldo\footnote{On leave from Dipartimento di Fisica,
Universit\`a degli Studi di Salerno, Baronissi, Salerno I-84081,
Italy.},
Steven M. Anlage}
\address{Center for Superconductivity Research, Department of Physics,
University of\\
Maryland, College Park, MD 20742-4111.}
\author{L. Maritato}
\address{Dipartimento di Fisica, Universit\`a degli Studi di Salerno,
Baronissi, Salerno I-84081, Italy.}
\date{\today}
\maketitle

\begin{abstract}
The surface impedance of Nb/CuMn (superconducting/spin-glass) bilayers has
been measured at $10$ $GHz$ with the parallel plate resonator technique
to obtain information about the exotic behavior of the order parameter in
superconducting/magnetic proximity systems. The data strongly differs from
the superconducting/normal-metal case, showing the magnetic nature of the
CuMn layer, which acts as a weak ferromagnet. The results are described in
the framework of two models for the electrodynamics of
superconducting/ferromagnetic (S/M) bilayers characterized by a
proximity-coupling length scale which is independent of temperature. 
\end{abstract}


\section{Introduction}

The issue of the interplay between superconductivity and ferromagnetism has
been an intriguing research opportunity for many years\cite{sm,htc}.
Exotic phenomena are predicted for
superconducting/ferromagnetic (S/M) coupled layered structures, such as
critical temperature oscillations versus the M layer thickness, or
spontaneous persistent currents in rings interrupted by an S/M/S junction 
\cite{russian,radovic,dem}. All of these properties depend on the presence
of a spatially dependent phase for the order parameter in the M layer
that, for suitable thicknesses, 
gives rise to a $\pi $ shift between adjacent superconducting
layers. From the experimental point of view, observations of $T_c$
oscillations, seen as an indirect proof of the $\pi $ phase, has been
reported for Nb/Gd multilayers\cite{NbGd1}, and Nb/Gd/Nb trilayers\cite
{NbGd2}, and also for Nb/CuMn (superconducting/spin-glass) multilayers\cite
{lucia1,lucia2}.

Among all the possible S/M proximity coupled layered structures, the
superconducting / spin-glass systems are very interesting. Because of the
weaker macroscopic magnetism, spin-glass systems offer a wider range of S
and M
thicknesses in which it is possible to study the influence of magnetism on
superconductivity, as compared to the superconducting/ferromagnetic case.
Moreover, when using a classical spin-glass such as CuMn, different coupling
regimes can be easily selected by changing the Mn concentration, a parameter
related to the effective exchange energy $I$.

A systematic study of the non-monotonic $T_c$ vs. $d_{CuMn}$ behavior in
Nb/CuMn multilayers has been previously performed in different coupling
regimes, by changing Mn concentration and relative thicknesses\cite{lucia2},
showing that an extension of the Radovic {\em et al.} theory for S/M
multilayers\cite{radovic} to the superconducting/spin-glass case seems very
plausible.

More insights into this problem may result from microwave surface
impedance
measurements, which have provided valuable information about
the inhomogeneous superconducting properties of layered systems\cite
{NbAl,NbCu}. There are dramatic differences in the electrodynamic
properties
of superconducting systems with non-uniform order parameter, compared to
the homogeneous case. And going backwards, analyzing the electrodynamic
properties provides information about the nature of the induced order
parameter in the M layers.

Here we present our results on the surface impedance of Nb/CuMn
bilayers with identical Nb base layers and different CuMn layer thicknesses.
In Sec. II we describe the sample fabrication and characterization. In Sec.
III we illustrate the experimental technique for the surface impedance
measurements and we show the data for the bilayers and compare them to the
bare Nb case and to previous results on conventional
superconductor/normal metal proximity coupled Nb/Cu bilayers\cite{NbCu}.
In Sec.
IV we present some theoretical background on the S/M proximity effect and
introduce two models for the electrodynamics of S/M bilayers, and in Sec.
V we apply the models to describe the penetration depth results. In Sec.
VI we
switch our attention to the surface resistance results, comparing them to
the theoretical behavior extracted by applying the models introduced in
Sec. IV, and finally in Sec. VII we summarize the results.

\section{Sample fabrication and characterization}

In this work we have focused on Nb/CuMn bilayers with a Mn concentration
of $ 2.7$\%. We have analyzed seven samples characterized by the same Nb layer
thickness ($d_S=1500$\AA ) and different CuMn layer thicknesses ($ 
d_M=30,60,90,120,150,180,240$\AA ), where the Nb is the first layer on the
substrate. The samples were grown all together in
the same deposition run on a 2 inch diameter Si(100) substrate (cut at the
end of the process) by a dual source magnetically enhanced dc triode
sputtering system with a movable substrate holder. The bilayers were
prepared in the same way as the multilayers studied 
previously\cite{lucia1,lucia2}. 

After the surface impedance measurements, the top CuMn layer of one pair
of samples (the one with $d_{CuMn}$=150\AA) was removed with a chemical
etching, by using a dilute $HNO_3$ solution, in order to characterize the
underlying Nb layer. First of all we have measured the surface impedance
of this single Nb layer with the technique described in the next section,
obtaining a
zero-temperature penetration depth of $\lambda (0) =1200$\AA. 
Then we patterned one of the films to perform a resistivity measurement
with the standard four-probe technique. The film has a resistive
$T_c=7.3K$\cite{tc}, a $10K$ resistivity $\rho_{Nb} (10K)=6\mu \Omega \cdot 
cm$, and a residual resistivity ratio $RRR=\rho_{Nb} (300K)/\rho_{Nb}
(10K)=1.9$. The low $RRR$ and the high residual resistivity are consistent
with both the observed low $T_c$ and high $\lambda (0)$ values
\cite{came,vaglio}, and are probably related to the presence of oxygen in
the Nb film. Similar values for all these parameters have been observed,
for example, in a Nb-O alloy with a 2\% oxygen content \cite{halbr}.
However, the low quality of our Nb does not compromise our work
because we compare the behavior of bilayers prepared with identical Nb base
layers, so that they differ only in the CuMn layer thickness. 
The $T_c$ values for the bilayers are slightly lower than the Nb film
(around $7.1K$), as confirmed by ac susceptibility measurements, and this
is one indication that the Nb and CuMn layers are proximity-coupled.

We have also prepared a series of CuMn single layer samples with the same
thicknesses as those appearing in the bilayers. We have performed Rutherford 
Backscattering Spectrometry (RBS) measurements on these samples
to get the actual Mn concentration and the actual thicknesses. Resistivity
measurements have also been performed on some of the thickest samples 
giving a $10K$ resistivity value of $\rho_{CuMn}(10K)=9\mu \Omega
\cdot cm$ for the $d_{CuMn}=240$\AA\ case, where this value is expected
to be a function of the CuMn layer thickness $d_{CuMn}$ 
(it increases upon reducing $d_{CuMn}$), due to finite-size
effects\cite{ken1}. Measurements on CuMn films of three different
thicknesses generally agree with this trend. The $\rho_{CuMn}
\left( T\right) $ curves for CuMn (figure 1) show an enhancement of the
resistivity around $130K$ often seen for spin-glass
materials\cite{mydosh}. Although this enhancement is not related to the spin
glass transition, it identifies the two regions where the transport 
electron - local spin
interaction is dominant ($T>130K$) and where the spin-spin interaction
begins to predominate ($T<130K$). The reduction of the resistivity while
decreasing the temperature below the peak value, together with the behavior
of other physical properties such as the specific heat, is widely
interpreted as an indication of the formation of local correlations and
mainly ferromagnetic clusters which gradually reduce the paramagnetic spin
disorder scattering\cite{mydosh,cam,myd}. This means that even above the
spin-glass freezing temperature $T_f$ the system cannot be described as a
simple paramagnet. Low-field ac susceptibility measurements of $T_f$ 
in bulk CuMn indicate $T_f^{bulk}=18K$ for a 2.7\%\ Mn
composition\cite{ford}, but a lower value is found for thin films
because of finite-size effects\cite{ken1,ken2,hoi}. A universal
dependence of the normalized freezing
temperature $T_f/T_f^{bulk}$ versus the CuMn layer thickness has been
observed while changing the Mn concentration\cite{ken1,ken2,hoi}.
Using the curve plotted in
ref.\cite{hoi}, we can estimate $T_f\simeq 14K$ for the thickest sample 
($d_{CuMn}=240$\AA) and lower values for the other samples until reaching 
$T_f\simeq 9K$ for $d_{CuMn}=30$\AA. Hence for all of our bilayers, the
freezing temperature is always larger than the superconducting critical
temperature. However, even in the case $T_f\le T_c$ no changes are
expected in the electrodynamic behavior\cite{spin-gl}.

\section{Surface impedance measurements}

Surface impedance measurements have been performed on the Nb/CuMn bilayers
and on the underlying Nb film at $10$ $GHz$ by using the parallel plate
resonator (PPR) technique \cite{ppr} with a $50\mu m$ thick Teflon
dielectric spacer.

\subsection{Experimental technique}

The resonator is formed from two nominally identical thin films clamped
face-to-face on a dielectric spacer. The sandwich is placed in a copper
chamber in thermal contact with a small
copper box in which liquid He can enter from the external dewar through a
needle valve. The system (chamber and He box) is enclosed in a vacuum can,
where we allow the presence of some He exchange gas ($P\sim 10\mu mHg$), 
to stabilize the temperature. By pumping on the
liquid He in the box the minimum temperature of the sample which can
be reached is $1.7K$. During the measurement the temperature is gradually
increased in discrete steps by means of a computer controlled heater which
is in thermal contact with the resonator enclosure.

Excitation of transverse electromagnetic (TEM) modes is accomplished by
using two $50\Omega $ microstrip antennas, whose position can be sensitively
varied by micrometers in order to optimize the coupling to the resonator.

For each temperature, we measure, as a function of frequency, the complex 
transmission coefficient $S_{21}$ (magnitude and phase) by using a vector
network analyzer (some measurements have been done with an HP8510C and
others with an HP8722D). The PPR\ resonant frequency $f_0$ and quality
factor $Q$ are extracted with
an inverse mapping fitting routine in the complex plane\cite{paul}.

The resistive loss of the superconducting films gives a contribution to
the measured $Q$ of
$Q_{res}=\pi \mu _0f_0d/R_S$, where $d$ is the spacer thickness and $R_S$
is the surface resistance of the films. There are, however, additional
extrinsic losses, such as the dielectric loss, which is independent of $d$,
and the radiation loss, which increases linearly with $d$\cite{ppr}. A
calibration can be performed by varying the dielectric spacer thickness, so
that all these factors can be uniquely determined. 
However, we found that the PPR modes sometimes lie too close to package
resonances,
thus changing the $Q$ values to varying extent depending upon the coupling
between the PPR and these parasitic modes\cite{gao}. Because of this
problem in measuring the intrinsic $Q$ values, we have only examined the
temperature dependence of $R_S$ and not its absolute value.

\subsection{Results}

Changes in the effective penetration depth have been extracted from the
measured resonant frequency by using the expression\cite{NbCu,thesis}:

\[
\Delta \lambda _{eff}\left( T\right) =\lambda _{eff}\left( T\right) -\lambda
_{eff}\left( T_0\right) =\frac d2\left[ \left( \frac{f_0\left( T_0\right)
}{ f_0\left( T\right) }\right) ^2-1\right] 
\]
where $T_0$ is the lowest temperature reached during the measurements
(usually $T_0\sim 1.7K$). 
In proximity-coupled systems $\lambda _{eff}$ is an overall
screening length which does not correspond to the individual screening
lengths of the constituents, because of the non-uniform nature of the
superconductivity in the bilayers.
Moreover, by using this expression we are not removing
the geometric correction due to the finite thickness of the
sample, given by the factor $coth(t/\lambda)$ for a homogeneous
superconductor, where $t$ is the film thickness \cite{klein}.
In our case, where $\lambda (0)$ is the order of $t$, this correction is
not negligible.
Therefore, even in the single-layer Nb film case, we are dealing with an
effective penetration depth.

Figure 2 shows the change in the effective penetration depth for the Nb
film
and the Nb/CuMn (S/M) bilayers, compared to previous results on Nb/Cu
(S/N) bilayers\cite
{NbCu}. Surprisingly the shape of the $\Delta \lambda _{eff}\left( T\right) $
curves for the S/M bilayers is not very different from the temperature
dependence for the Nb, but shows only an enhancement, while a strong
linear-in-temperature character was evident in $\Delta\lambda_{eff}(T)$
for Nb/Cu. Moreover there is no
systematic dependence of $\Delta\lambda_{eff}(T)$ on the CuMn layer
thickness, in striking contrast to the case of Nb/Cu bilayers, where a
strong dependence of $\Delta \lambda _{eff}\left( T\right) $ on the normal
layer thickness has been observed. Results similar to Nb/Cu have also been
obtained for Nb/Al bilayers\cite{NbAl}.

Figure 3 shows the temperature dependence of the effective surface
resistance, minus the residual value at $T_0$, for the Nb film and the
bilayers. Again the Nb/Cu data\cite{NbCu}, corrected for extrinsic losses, 
have been plotted for comparison. The data have
been extracted from the measured quality factor $Q$ by using the relation
$ R_S=\pi \mu _0f_0d/Q$\cite{ppr}, neglecting dielectric and radiation
losses\cite{NbCu}, and then the low temperature residual resistance
$R_{S0}=R_S\left( T_0\right) $ has been subtracted. From
this point of view it is interesting to observe that as for $\Delta \lambda
_{eff}\left( T\right) $, also the $R_S\left( T\right) $ behavior is  
similar to the BCS behavior (Nb data are shown as inverted triangles on
fig. 3), and
again there is no systematic variation with $d_{CuMn}$. These results are 
in striking  contrast to the results on Nb/Cu, where also a
low-temperature downturn of $R_S\left(T\right) $ was observed that is
missing here. Note, also, that the Nb/CuMn data fall into two groups: one
composed of the thin CuMn layer films, which are very close to the pure Nb
film, and a second group composed of thicker films (together with the
$90$\AA\ CuMn sample) which show enhanced $R_S-R_{S0}$.

In figures 2 and 3 the changes in the effective penetration depth and the
effective surface resistance for the Nb underlayer are reported together
with the results for the bilayers. In figure 4 these quantities (solid
symbols) are compared to the corresponding intrinsic one (open symbols), 
obtained by performing a finite thickness correction, as given by Klein
{\em et. al.}\cite{klein}, in conjunction with the BCS-M\"{u}hlschlegel
\cite{muh} fit. The solid line in the figure is the BCS-M\"{u}hlschlegel 
fit to the intrinsic data, which gives $T_c=7.7K$
and $\lambda (0)=1200$\AA. In both cases the effective quantity is strongly
enhanced compared to the intrinsic one, since $\lambda (0) \sim t$.
Concerning $R_S$, again we are neglecting the dielectric and radiation
losses, which partly explains the high residual value shown in figure 4.

\section{Models for the S/M bilayer electrodynamics}

To understand the origin of the different behavior in S/M and S/N proximity
systems we need to analyze the order parameter that describes the extent of
the proximity coupling, given by the pair potential $\Delta \left( 
\overrightarrow{r}\right) =V\left( \overrightarrow{r}\right) \left\langle
\psi _{\uparrow }\left( \overrightarrow{r}\right) \psi _{\downarrow }\left( 
\overrightarrow{r}\right) \right\rangle $. Here $V\left(
\overrightarrow{r} \right) $ is the electron-electron 
attractive interaction responsible for 
superconductivity and $\left\langle \psi _{\uparrow }\left( 
\overrightarrow{r}\right) \psi _{\downarrow }\left( \overrightarrow{r} 
\right) \right\rangle $ is the probability amplitude to find a Cooper pair
in the position $\overrightarrow{r}$. In the usual picture of the single
frequency approximation in S/N bilayers $\Delta \left( \overrightarrow{r} 
\right) $ decreases from its bulk value in the vicinity of the S/N interface
in the S material, while a non-zero order parameter is induced on the N
side, decaying exponentially as the free surface is approached\cite{deg}.
For the penetration depth, the widely accepted approximation\cite{deut} $ 
\lambda \left( \overrightarrow{r}\right) \varpropto \frac
{\displaystyle 1}{\Delta \left(\overrightarrow{r}\right) }$ gives it an
exponential dependence in the N (non-magnetic) layer.

The magnetic case is more complicated and a new treatment is required. 
Based on our previous results for $T_c$ vs. $d_{CuMn}$ 
oscillations\cite{lucia1,lucia2}, we want to apply the same Radovic {\em \
et al}. theory\cite{radovic} to propose a
spatial dependence for $\lambda \left( \overrightarrow{r}\right) $.
Radovic {\em \ et al.} show that in a magnetic metal,
where $\Delta _M=0$ because the BCS coupling is identically zero, the Green
function $F_M$ describing the condensate of pairs is non-zero, due to the
proximity of S. Its real part (and imaginary part) exhibits an oscillatory
behavior damped by the exponential decay usual in S/N systems. Moreover, in
this theory, developed in the dirty limit, the characteristic penetration
length of the Cooper pairs in M, defined as $\xi _M=\sqrt{4\hbar D_M/\left|
I\right| }$, depends on the diffusion coefficient $D_M=\frac 13v_Fl_M$ and
on the exchange energy $I$ but is temperature independent and typically
much smaller
than the corresponding length in a normal metal with $D_N=D_M$, that is
given by $\xi _N=\sqrt{\hbar D_N/2\pi k_BT}$. Here $v_F$ is the Fermi
velocity and $l_M$ is the mean free path of quasiparticles in the M layer.
A temperature independent and small penetration of Cooper pairs in M was
already predicted by the combination of the de Gennes-Werthamer theory
for S/N proximity effect\cite{deg,wer} and the Abrikosov and Gor'kov
analysis
for the role of paramagnetic impurities in a superconductor\cite{gor}, as
shown in ref.\cite{htw}.

Even if $\Delta _M=0$, due to the proximity effect the Cooper pair
probability amplitude $\left\langle \psi _{\uparrow }\left(
\overrightarrow{r }\right) \psi _{\downarrow }\left( \overrightarrow{r}
\right) \right\rangle _M
$ is non-zero and has the same oscillating and damped behavior of $F_M$. If
we call $x$ the axis perpendicular to the interface and we identify the
interface position as $x=0$, so that the M layer extends from $x=0$ to $ 
x=-d_M$ while the S layer extends from $x=0$ to $x=d_S$ (see inset of fig.
5), in the bilayer case this amplitude can be written as:

\[
\left\langle \psi _{\uparrow }\left( x\right) \psi _{\downarrow }\left(
x\right) \right\rangle _M  \varpropto \exp \left( k_Mx\right) \hspace{1in}
-d_M\le x\le 0
\]
with $k_M$ complex\cite{km}, which means that the Cooper pairs acquire a 
spatially dependent
phase in the magnetic layer, while their density, which is proportional to
$ \left| \left\langle \psi _{\uparrow }\left( x\right) \psi _{\downarrow
}\left( x\right) \right\rangle _M\right| ^2$, decays exponentially. In this
picture, assuming now $\lambda _M\left( x\right) \sim \left| \left\langle
\psi _{\uparrow }\left( x\right) \psi _{\downarrow }\left( x\right)
\right\rangle _M\right| ^{-1}$ (a local inverse proportionality between
$\lambda_M^2(x)$ and the Cooper pair probability density) we have again an
exponential dependence for $ \lambda _M$, but with a characteristic length
$\left({\rm Re}\, k_M\right) ^{-1}\sim \xi _M$ that is 
temperature-independent and smaller than that observed in the S/N case:

\[
\lambda _M\left( x,T\right) =\lambda _M\left( 0,T\right) \exp \left[ -( 
{\rm Re}\,k_M)x\right] \hspace{1in} -d_M \le x \le 0
\]
All the temperature dependence of $\lambda _M$ enters in the coefficient $
\lambda _M\left( 0,T\right) $ for which the BCS temperature dependence
can be assumed.

Therefore, on the basis of this simple theory, we conclude that the
oscillating behavior of the induced pair amplitude cannot be detected in
our experiment, since
we are only sensitive to its modulus. However we can check to see if our
experimental results are consistent with a picture in which $\lambda
_M\left( x,T\right) $ has a strong exponential behavior with a
characteristic length which is not dependent on temperature. This also
means that we can use existing models for the electrodynamics of S/N
bilayer systems\cite{thesis,pamb} with minor changes.

We shall now summarize the two leading models of the electrodynamics of
S/N bilayers. In the first model (Model I) we assume that the
non-superconducting layer is active in screening the applied magnetic field
and, on the other side, the S layer is unaffected by the presence of the
non-superconducting layer, so that $\lambda _S$ is uniform across all the
S layer and its
temperature dependence is that given by the BCS theory\cite{muh}. In the M
layer we take $\lambda _M\left( x,T\right) =\lambda _M\left( 0,T\right) \exp
\left( kx\right) $ (see inset of fig. 5). Because of the small value we
expect for $k^{\_1}$ in
the magnetic case, which means almost no screening of the magnetic field by
the M layers, we also consider a second model (Model II) in which the M
layer does no screening at all $\left( \lambda _M\left( x\right) \rightarrow
\infty \right) $\cite{infinity}, while the superconducting properties of
the S layer are suppressed near the interface. Following de
Gennes\cite{deg2}, in Model II we assume in particular that the order
parameter in S decreases, upon approaching
the interface, as $\Delta _S\left( x,T\right) =\Delta _0\tanh \frac{x-x_0}{
\sqrt{2}\xi _S\left( T\right) }$ ($x>0$), which means that the penetration
depth is enhanced following the dependence $\lambda _S\left( x,T\right)
=\lambda_{S0}\left( T\right) \coth \frac{x-x_0}{\sqrt{2}\xi _S\left( T\right) 
}$ ($x>0$),
where $\lambda _{S0}\left( T\right) $ is the bulk penetration depth,
$\xi_S$ is the superconducting coherence length, $x_0=-
\frac{\xi _S}{\sqrt{2}}\ln \left( \frac{\sqrt{2}b}{\xi _S}+\sqrt{\frac{2b^2}{
\xi _S^2}+1}\right) $ and $b$ is the extrapolation length, proportional to
the coherence length of the adjacent non-superconducting layer (see inset 
of fig. 5). 

We consider Models I and II to be two extreme cases in which the altered
screening in either the M or S layer dominates the proximity screening.
Actually in general the real process involves both kinds of physics,
although we expect it to be dominated by one or the other. Moreover, in
both of the models we are not explicitly taking into account the
transparency of the S/M interface for Cooper pairs, that is an important
parameter in these systems\cite{aarts}; however, the behavior at the
interface is hidden in the parameter values.

The details about the calculations of the tangential magnetic
field $H\left( x\right) $, the supercurrent density $J_S\left(
x\right) $ and the effective penetration depth $\lambda _{eff}$
may be found in refs.\cite{thesis,pamb}. These expressions have been used,
with appropriate modifications for the S/M case, in the discussion below.

\section{Comparison of data and models}

We found previously that in the S/N bilayer case, Model II could
not describe the $\Delta \lambda
_{eff}\left( T\right) $ data, because the theoretical behavior is not
very different from a BCS s-wave temperature dependence\cite{thesis,pamb}.  
Model I, instead, was
found to describe the $\Delta \lambda _{eff}\left( T\right) $ data very well
\cite{NbAl,NbCu}. Here we report the analysis of the Nb/CuMn data with
both of the models introduced in the previous section.
Actually in the S/M case we do not expect much screening in the M layer,
in which case Model II may be more appropriate. However, in principle we
do not know {\em a priori} if CuMn is really acting as a ferromagnet in
the screening process. And through Model I we can do a more direct
comparison with the S/N case, in particular by looking at the proximity
length scale $k^{-1}$ value and its temperature dependence.

\subsection{Model I}

In the S/N case, it was crucial to use a proximity length
scale $k^{-1}$ which was temperature dependent to get a good fit to the
data. In this way in the low $T$ region $\lambda _{eff}(T)$ is a strongly
increasing function of temperature and the shape of the curves depends
on $d_N$. In particular, $k^{-1}\left( T\right) \sim T^{-1/2}$ was found
for Nb/Al\cite{NbAl} and $k^{-1}\left( T\right) \sim T^{-2}$ for
Nb/Cu\cite{NbCu}. In contrast, with 
$k$ independent of $T$, $\lambda _{eff}\left( T\right) $ is
flat at low $T$ (below $T/T_c\simeq 1/4$, with the $k^{-1}$ value used here,
that is reported below) independent of $d_M$, and $\Delta \lambda
_{eff}\left( T\right) $ is essentially zero for all the
non-superconducting layer thicknesses.
At higher temperatures the curves increase together in a manner which is
independent of $d_M$. This is just the temperature behavior we have observed
in the S/M case. Thus the Nb/CuMn experimental results are qualitatively 
consistent with a $\xi _M$ which is independent of temperature.

The solid curve in figure 5, which describes well all the $\Delta
\lambda_{eff} (T)$ data on the Nb/CuMn bilayers, has been obtained using
reasonable values for all the parameters, as shown in Table 1.
However, even though the agreement with the model looks good, Model I is
not really appropriate to describe S/M bilayers, where the proximity
length scale $k^{-1}$ is very short ($k^{-1}<<\lambda_M(0,0)$)\cite{deut}.
Indeed, in Model I the assumption $\lambda _M\left( x\right) \sim \left|
\left\langle \psi _{\uparrow }\left( x\right) \psi _{\downarrow }\left(
x\right) \right\rangle _M\right| ^{-1}$ forces the screening length to be
regulated by the proximity length scale $k^{-1}$. In the S/M case this
implies huge current densities in a narrow region ($\sim \xi_M$) near the
interface , while there is a small screening activity in the S layer,
which is not a physically reasonable scenario.

In summary, using Model I, we found a temperature independent proximity
effect correlation length (as expected on the basis of the Radovic {\em et
al.} picture for S/M multilayers\cite{radovic}), and we learned that
almost all the screening activity happens in the superconducting layer. 
These results confirm that the CuMn layer is acting as a ferromagnet in
the screening process.
However, the screening activity picture given by Model I is unphysical.
The main reason is that the spatial variation of $\lambda_S$ in the
superconductor is not taken into account and this can be a good
approximation in the S/N case but not in the S/M case. These problems are
addressed in Model II.

\subsection{Model II}

In Model II we are assuming that the penetration depth is infinitely large
in M and a decreasing function of $x$ in S going from the interface ($x=0$)
to the opposite edge ($x=d_S$), and the extrapolation length $b\sim \xi _M$
dictates the behavior at the interface (see inset of figure 5). As with
$k^{-1}$ in Model I, this parameter ($b$) is temperature independent in
the S/M case. As in Model I we again do not obtain any dependence of
$\Delta \lambda_{eff} (T)$ on the M layer thickness. 

The parameters in Model II are the Nb critical temperature $T_c$, the
extrapolation length $b$, the zero-temperature superconducting coherence
length $\xi _S(0)$, and the zero temperature penetration depth $\lambda _{S0}
$ far from the interface. In Model II only parameters characteristic of
the S layer appear, except for $b$, which is the only parameter directly 
dependent upon the nature of the non superconducting layer. 
We have used $T_c=7.7K$ and $\lambda _{S0}=1200$\AA, as obtained from the
analysis of the single layer Nb film, $b=\xi_M=19$\AA, which comes from
the previous work on the $T_c$ oscillations vs. $d_{CuMn}$
multilayers\cite{lucia2} (however, a value between 0 and 100 \AA\ still
describes the data well), and $\xi _S\left( 0\right) $=150\AA\ (Table 1). 
The $\xi _S(0)$ value has been chosen as close as possible to the dirty
limit coherence length $\xi _S=\sqrt{\hslash D_S/2\pi k_BT_c}$
\cite{radovic}, with $D_S=v_F l_S/3$ the diffusion
coefficient in S and $l_S$ the mean-free path in S. From the measured low
temperature resistivity value we get $l_S=63$\AA , which gives
$\xi_S=93$\AA. The curve for the Model II prediction of
$\Delta\lambda_{eff} (T)$ is shown in figure 5 (dashed 
line). This model does an excellent job of fitting the data, and
moreover addresses the problems encountered in Model I. In particular it
takes into account the suppression of the superconducting properties in
the S layer at the interface, which is not negligible in the S/M case.
Indeed, due to the very small $b$ value ($b \sim$ 0 to 100\AA), we
observed here $\Delta_S(x=0)/ \Delta_{S, bulk} \sim $ 0 to 0.4 for $T
\rightarrow 0$ for all the Nb/CuMn bilayers (in particular for $b=19$\AA\
this ratio is $\sim 0.09$), where this ratio is even smaller at higher
temperatures.

In conclusion Model II captures the essence of the 
electrodynamics of S/M bilayers (screening dominated by the S layer
which has a strongly suppressed order parameter near the interface) and
the fit is done with independently determined parameters.

\section{Surface resistance in Model II}

In this section we want to extend Model II to describe also the surface
resistance data. The surface resistance $R_S$ can be calculated by using
the relation\cite{NbAl,thesis}:

\[
R_S\left( T\right) =\frac{\mu _0^2}{H_0^2 }\int
\limits_{-d_M}^{d_S}\frac{\sigma _1\left( x,T\right) }{\sigma _1^2\left(
x,T\right) +\sigma _2^2\left( x,T\right) }J^2\left( x,T\right) dx 
\]
where $\sigma _1\left( x,T\right) $ and $\sigma _2\left( x,T\right) $ are
the real and imaginary parts of the local conductivity, $H_0$ is the
applied field and $J$ is the total current density.
Again, we have to remember that we are dealing with an effective surface
resistance, not only because of the non-uniformity of the magnetic
screening, but also because of the finite sample thickness.

To obtain an estimate for $R_S(T)$ in the bilayers, in a first
approximation we have used the current density calculated before with
Model II to get $\lambda _{eff}$, even though it was
derived neglecting the normal currents.
As for the local conductivity, we have taken $\sigma _2\left( x,T\right)
=\left[ \omega \mu _0\lambda ^2\left( x,T\right) \right] ^{-1}$, where we
have used the Model II spatial dependence of
$\lambda(x,T)$, while for $ \sigma _1\left( x,T\right) $ we have used a
generalized Mattis-Bardeen\cite{MB} expression in which the local BCS gap
$\Delta_S \left( x,T\right)$ used in model II (the expression is reported
in Sec. IV) replaces the spatially uniform one found in
homogeneous superconductors\cite{thesis}. 

Model II only calculates the contribution to $R_s$ coming from S;
the contribution from M is only an additive constant term.
Therefore we obtain for $R_S(T)$ something that goes to zero at low
temperature, so to compare with the experimental behavior, we subtracted 
the lowest temperature residual resistance $R_{S0}=R_S(T_0)$ from the data
(as in figures 3 and 6). 

The theoretical curves in figure 6 have been obtained with the same 
parameter values used for the $\Delta \lambda _{eff}\left( T\right)$
analysis (i.e. $T_c$, $\lambda _{S0}$, $\xi _S\left( 0\right)$, $b$). 
The gap value has been chosen equal to the BCS value $\Delta
_0=1.76k_BT_c$, and a complete freedom has been left for the low temperature
normal state conductivity $\sigma_S$ in S. The $\sigma_S$ values found
with the fit procedure are only slightly lower than that measured. Table 1 
summarizes the parameter values, and the fits are shown as dashed lines in
figure 6.

In figure 6 we also show fits obtained using Model I (solid lines). In
this case, to evaluate the generalized Mattis-Bardeen $\sigma_1(x,T)$ in
M, for simplicity we did not take into account any effect due to the
spatial dependence of the phase of the order parameter in M and we used a
decaying real exponential dependence for $\Delta_M (x,T)$, as for the S/N
bilayer case\cite{rs}. 
With Model I unreasonably large $\sigma_S$ values had to be used to get
close to the data (the fit parameters are reported in Table 1).

In summary, the theoretical treatment of the surface resistance with Model
II gives a satisfactory understanding of the $R_S(T)$
behavior using the same parameter values as for the analysis of the
penetration depth temperature and $d_{CuMn}$ dependence data.

\section{Conclusions}

In conclusion, microwave surface impedance measurements have been performed
on Nb/CuMn superconducting/spin-glass proximity coupled bilayers. Both
$\Delta \lambda _{eff}\left( T\right) $ and $R_S\left( T\right) $ results
are very different from the Nb/Cu case, showing
that the superconducting properties of the Nb layer are strongly suppressed
near the interface and the CuMn layer does not participate too much in the
screening of the applied rf magnetic field. These are exactly the results we
expect for a proximity effect between a superconductor and a ferromagnet.
Therefore they confirm that the CuMn layer is acting as a weak ferromagnet.
Moreover, these measurements can be described with a proximity effect
correlation length which does not depend on temperature, consistent with
previous data on $T_c$ oscillations vs. CuMn layer thickness, and the
Radovic {\em et al.} theoretical picture for S/M systems\cite{radovic}.
This work sets also the stage to investigate more directly the presence of
a $\pi$-phase shift in superconducting/ferromagnetic (spin-glass) layered
structures with the Nb/CuMn system.

\section{Acknowledgments}

We thank M. S. Pambianchi for his contributions in the design of the probe
and in the development of model calculations for the S/N case, P. Fournier
for assistance with the resistivity measurements, and R. P.
Sharma and S. Choopun who performed the RBS analysis.

This work is supported by the National Science Foundation through grant
number DMR-96-24021.

\begin{figure*}
\caption{Resistivity versus temperature curve for the $150$\AA\ thick
CuMn sample. Similar behavior has been observed for CuMn samples of
different thickness.}
\vspace{0.5in}
\caption{Changes in the effective penetration depth, $\Delta\lambda_{eff}
(T)$, with respect to temperature for the bare Nb film
and the Nb/CuMn bilayers (open symbols and left axis) compared to the
Nb/Cu data (solid symbols and right axis). The Nb/Cu data are shown with
an arbitrary offset in the vertical direction for clarity. The data are
plotted versus the normalized temperature $T/T_c$, where the critical
temperature used here for our data ($6.6K$) is lower than the
resistive value measured for the Nb underlayer ($7.3K$). The Cu layer
thicknesses in \AA\ are reported in the legend, while the CuMn layer
thicknesses appear in figure 6 next to the symbols used in the figure.} 
\vspace{0.5in}
\caption{Effective surface resistance ($R_S$) minus the residual value at
the minimum temperature ($R_{S0}$) vs. the normalized temperature $T/T_c$
for the Nb film (inverted triangles) and the Nb/CuMn bilayers, together
with the $R_S (T)$ data
for the Nb/Cu case corrected for extrinsic losses. We have adopted the
same symbols as in figure 2. For the CuMn layer thicknesses refer to fig.
6 and for the Cu layer thicknesses to fig. 2.} 
\vspace{0.5in}
\caption{Effective (solid symbols) and intrinsic (open symbols) changes in
the penetration depth (circles and left axis) and surface resistance
(squares and right axis) of the Nb undelayer. The effective quantities
have been altered by geometrical effects due to the finite thickness of
the films. The solid line in the figure is the fit to the BCS theory.}
\vspace{0.5in}
\caption{Comparison between the experimental changes in the effective
penetration depth for the Nb/CuMn bilayers and the theoretical curves
obtained with the models described in the text. The CuMn layer thicknesses
in \AA\ appear in fig. 6 (we are adopting the same symbols). Inset:
S/M bilayer geometry with the
magnetic-field boundary conditions and the schematic variation of order
parameter and penetration depth in Model I and Model II (in Model II the M
layer does not screening at all, i. e. $\lambda _M\left( x\right)
\rightarrow \infty$).} 
\vspace{0.5in}
\caption{Surface resistance minus the residual value at the minimum
temperature for all the bilayers together with the theoretical curves
obtained with the models presented in the text. The theoretical curves
describing the lower $R_S-R_{S0}$ data group are characterized by a lower
Nb $10K$ conductivity value ($\sigma_S$). The legend shows the CuMn layer
thicknesses in \AA.}
\end{figure*}

\vspace{8.0in} 
Table 1. Summary of the values of all the parameters
appearing in this work. The fitting parameter values obtained in the
$\Delta \lambda_{eff} (T)$ and $R_S(T)$ analyses in both Model I and
Model II are compared with the corresponding measured or otherwise
determined values. In particular $T_c$, $\sigma _S$ and $\sigma _M$ have
been resistively measured, $\lambda_S(0)$ was determined by fitting the
frequency shift data for the bare Nb to the BCS theory (in parentheses we
show also the $T_c$ obtained with this procedure), $\xi
_S\left(0\right)$ is the predicted dirty limit value with the
BCS-determined $T_c$, and for $k^{-1}$ and $b$ we report here the $\xi_M$
value obtained in the previous work on the $T_c$ oscillations in Nb/CuMn
multilayers with the same Mn concentration\cite{lucia2}.

\vspace{.5in}

\begin{tabular}{|l|c|cc|cc|}
\hline \hline

parameter & 'measured' & from $\Delta \lambda _{eff}\left( T\right)$ fit & 
& from $R_S\left( T\right) $ fit &  \\ 
&  & Model I & Model II & Model I & Model II \\ 
\hline

$T_c\left( K\right) $ & $7.3$ ($7.7$) & $7.7$ & $7.7$ & $7.7$ & $7.7
$ \\ 
$k^{-1}\left( \text{\AA }\right) $ & $\sim \xi _M=19$ & $19$ 
& - & $ 19 $ & - \\ 
$b\left( \text{\AA }\right) $ & $\sim \xi _M=19$ & - &  0 to 100  & -
& 0 to 100\\
$\xi _S\left(0\right) \left( \text{\AA }\right) $ & $93$ & -
& $150$ & - & 150 \\ 
$\lambda _S\left(0\right) \left( \text{\AA }\right) $ & $1200$
& $1200$ & $1200 $ & $1200$ & $1200$ \\ 
$\lambda _M\left( 0,0\right) \left( \text{\AA }\right) $ & - & $\sim 120$
& - & $\sim 120$ & - \\ 
$\sigma _S\left(10^7/ \Omega \cdot m\right)$ & $1.7$
& - & - &  5 to 10 & .7 to 1.2 \\ 
$\sigma _M\left(10^7/ \Omega \cdot m\right)$ & $1$ & - & - & $1$ & -\\ 
\hline\hline
\end{tabular}


\begin{references}
\bibitem{tc} The value reported in the text for $T_c$ is the
zero-resistance transition temperature, while the onset is at 7.5K and the
fit of the penetration depth data to the BCS theory gives $T_c=7.7K$.
\bibitem{spin-gl} Sharp features are observed at the freezing
temperature in ac susceptibility, magnetization, anomalous Hall
effect, M\"{o}ssbauer effect, $\mu^+$ depolarization, while in specific
heat, resistivity, thermopower and electron spin resonance, only broad
changes over a wide temperature range have been observed\cite{mydosh}.
\bibitem{km} The general expression for $k_M$ is $k_M^2=2\frac{\hbar
\omega +iI}{\hbar D_M}sgn\omega $, where $\omega =\frac{\pi k_BT}\hbar
\left( 2n+1\right) $ are the Matsubara
frequencies, but the expression $k_M^{-1}=\left( 1-i\,sgn\omega
\,sgnI\right) \sqrt{\frac{\hbar D_M}{4I}}$, valid in the limit
$I>>k_BT_{cs}$, is generally used\cite{radovic}.
\bibitem{infinity} To be more precise, in Model II we should assume that
$\lambda_M$ approaches the normal metal skin depth $\delta$. However, the
CuMn layer thickness is much smaller than $\delta$ and there is no
screening.
\bibitem{rs}  As pointed out in Sec. IV, $\Delta_M=0$ in M layers,
therefore, to be precise, we should do a treatment in terms of the Green
function $F_M$ or of the pair amplitude $\left\langle \psi _{\uparrow}
\psi _{\downarrow } \right\rangle _M$.
\end{references}
\end{document}